\documentstyle[preprint,aps,tighten,epsfig]{revtex}

\begin{document}
\draft

\title{Extended van Royen-Weisskopf formalism for lepton-antilepton meson decay
widths within non-relativistic quark models}

\author{L.A. Blanco$^{(1)}$, R. Bonnaz$^{(2)}$, B. Silvestre-Brac$^{(2)}$,
\\F. Fern\'andez$^{(1)}$, A. Valcarce$^{(1,3)}$}
\address{
(1) Grupo de F\' \i sica Nuclear,Universidad de Salamanca,\\
E-37008 Salamanca, Spain. \\
(2) Institut des Sciences Nucl\'eaires, 53 Av. des Martyrs,\\
F-38026 Grenoble Cedex, France. \\
(3) Departamento de F\'{\i}sica Te\'orica, Universidad de Valencia,\\
E-46100 Burjassot, Valencia, Spain.}

\maketitle

\date{today}

\begin{abstract}
The classical van Royen-Weisskopf formula for the decay width of a meson
into a lepton-antilepton pair is modified in order to include non-zero quark
momentum contributions within the meson as well as relativistic effects.
Besides, a phenomenological electromagnetic density for quarks is
introduced. The meson wave functions are obtained from two different models:
a chiral constituent quark model and a quark potential model including
instanton effects. The modified van Royen-Weisskopf formula is found to
improve systematically the results for the widths, giving an overall good
description of all known decays.

\vspace{1cm} \noindent PACS: 12.39.Jh, 14.40.-n, 13.20.-v, 12.20.Ds

\vspace{1cm} \noindent Keywords: Nonrelativistic quark model; Meson leptonic
decays
\end{abstract}

\section{Introduction}

In a previous paper \cite{bo-bl} we studied meson strong decays in the
framework of non-relativistic quark models. In that work, a comparison was
made between a potential including only gluon exchanges and a potential
based on a chiral constituent quark model ($\chi$QM). Although originally
designed for the nucleon-nucleon ($NN$) interaction, the $\chi$QM has been
successfully applied to the baryon and meson spectra, besides the meson
strong decays. This is a stringent test for the model, as there are only a
few free parameters, which are taken to be the same for the $NN$ interaction
and the hadronic properties. Such a unified description reveals also the
capability of a non-relativistic approach to take care of a wide set of
phenomena, even in situations where it seems not justified to apply it.

The calculation of the spectrum is just an average of the meson wave
function; therefore, to discriminate between different models it is
necessary to rely on more sensitive observables. Meson strong decays were
already calculated and, as already said, with rather good results. But
electromagnetic transitions are even more interesting, since the transition
operator is perfectly known and therefore they constitute a very drastic
test of the wave functions. The present work is a first step to the study of
electromagnetic properties, it deals with the decay of a meson into a
lepton-antilepton pair.

Since many years the study of the decay width of a meson into a
lepton-antilepton pair has been done in the frame of the van Royen-Weisskopf
formula \cite{royen}. This work is based on a second order Feynman graph,
where the quark-antiquark pair which forms the meson annihilates into a
virtual photon which later on produces the lepton-antilepton pair. This
implies that only neutral mesons with $J^{P}=1^{-}$ are allowed to decay
within this scheme. In practice, this concerns $^{3}S_{1}$ and $^{3}D_{1}$
resonances (strictly speaking a true state can be a mixing of both, due to
the tensor force, but in this study we discard it so that the corresponding
states are decoupled). The calculation is realized through the familiar
Dirac trace method in order to evaluate the squared modulus of the
transition amplitude. In the original work \cite{royen}, besides a missing
factor 3 due to the ignorance of colour degrees of freedom at that time,
there are mainly two approximations: first of all to consider point-like (bare) charged
quarks, and secondly to consider the quarks having
zero momentum inside the mesons.

The approximation of point-like charged quarks is not consistent with 
the assumptions of the constituent quark model. Constituent
quark masses arise when an initially massless quark moves
through the
QCD vacuum. Due to the strong $qq$ interaction, the quarks get dressed
with a polarization cloud of $q\bar{q}$ pairs. Therefore,
a finite electromagnetic size appears due to the quark-antiquark
dressing.

The second approximation is the responsible for the leptonic decay widths being 
determined by the value of the meson wave function at the origin.
This quantity is very sensitive to the underlying potential, so that meson
leptonic decay widths are really a very important observable to discriminate
between different wave functions. Within the original van Royen-Weisskopf
formalism, $^{3}D_{1}$ states, having a null wave function at the origin,
are not allowed to decay. Nevertheless, some decays of such states have been
observed experimentally, indicating limitations to the applicability of the
van Royen-Weisskopf formula.

There have been other studies of the leptonic decay widths of mesons. In
Ref. \cite{mar} the leptonic widths of light vector mesons were calculated
in a modified MIT bag model under the assumption that the total
quark-antiquark momentum is zero in the center-of-mass system of the meson.
Ref. \cite{bar} used a similar formalism but based on the dominance of a
scalar-vector harmonic confining interaction for the mesons. Both
calculations neglected residual interactions like one-gluon exchange.
Extensions of the van Royen-Weisskopf formalism can also be found in the
literature. The Bonn group \cite{bey} obtained a quasi-relativistic formula
by considering $p/m$ terms in the decays, i.e., without neglecting the small
components of the Dirac spinors of quarks. The modification to the van
Royen-Weisskopf results for $S$-waves ranges between 5 and 15 \% depending
on the state. A non-zero, although very small, probability for $D$-wave
meson decays was obtained. Ref. \cite{lou} also considered relativistic
corrections to the van Royen-Weisskopf formula in a covariant formulation of
light-front dynamics. The difference with respect to non-relativistic
calculations can be as large as 60 \% in the charm sector, being smaller ($%
\sim $ 30 \%) for bottomonium.

Our purpose in this work is to extend the van Royen-Weisskopf formula in
order to take care of the {\it a priori} important aspects mentioned
above. First of all, to consider the momentum distribution of quarks inside a
meson,
as well as kinematical relativistic effects in the
non-relativistic wave function. Secondly, to study the role
played by the constituent quark
electromagnetic size. We will work in the framework of the constituent quark
model of Ref. \cite{jpg} assuming that the meson decay proceeds via the $q
\bar{q}$ annihilation as in Ref. \cite{royen}. Therefore, our formalism will
be on the same spirit as that of Ref. \cite{bey}. In order to calculate the
corresponding Feynman amplitude we modify our non-relativistic wave function
including a relativistic spinor. Finally, we want to test the sensitivity of
the results to the corresponding non-relativistic wave functions. For this
purpose we will perform an exhaustive study of all seen transitions using
the two different quark-antiquark potentials previously mentioned.

The paper is organized as follows. In the next section we calculate the
decay width of the transition taking into account all the improvements
presented above. The main features of the quark-antiquark potentials are
refreshed in section \ref{sec-mod}. The results are presented and analyzed
in section \ref{sec-res}. In the last section we summarize the conclusions
of our work. Some technical aspects as well as the specific form of the
electromagnetic size for the quarks
are detailed in the appendices.

\section{Calculation of the decay width}

\label{sec-calc}

\subsection{Notation}

As the formalism is very involved let us first of all make explicit the
notation we will use.

For the meson: mass $M$, momentum in the center-of-mass reference frame $%
\vec{P}=0$, total angular momentum $J$ and projection on a fixed $z$-axis $%
\mu_J$, orbital angular momentum $L$ and projection $\mu_L$, spin $S$ and
projection $\mu_S$, wave function in coordinate space $\Psi(\vec{r}) = R(r)
Y_{L \mu_L} (\theta, \varphi)$, and wave function in momentum space $\phi(%
\vec{p}) = R_p(p) Y_{L \mu_L} (\theta, \varphi)$ (a more correct expression
for the arguments of the last spherical harmonics would be $(\theta_p,
\varphi_p)$, however for typographical facility we still use $(\theta,
\varphi)$ when no confusion arises).

For the quark: mass $m_q$, charge $e_q$ (in units of the elementary positive
charge $e$), momentum $\vec{p}$, kinetic energy $E_p=\sqrt{m_q^2+p^2}$, spin
wave function $\vert \Phi_{M} \rangle $ with projection $M$, and helicity
wave function $\vert \chi_{\lambda} \rangle $ with projection $\lambda$.

For the lepton: mass $m_l$, charge $-e$ , momentum $\vec{q}$, kinetic energy 
$E_q=\sqrt{m_l^2+q^2}$ (do not confuse this $q$ index with the index
relative to the quark; in principle the context is not ambiguous) and spin
projection $\xi$.

For the antiparticle (antiquark and antilepton): the mass is the same as the
one of the particle, the charge is opposite and, since we are in the rest
frame, the momentum is opposite and the kinetic energy identical.

In addition, the virtual photon four-momentum $p_\gamma$ verifies $%
p_\gamma^2=M^2$ by energy conservation, and we express the Clebsch-Gordan
coefficients as $\langle j_1 m_1 j_2 m_2 \vert j_3 m_3 \rangle$.

\subsection{Leptonic decay width}

We calculate the decay width in the $S$ matrix formalism. One should
remember that the Feynman diagram refers to free particles with given
momenta; therefore, in a meson, we must take care of the fact that the
fundamental amplitude for the quarks must be integrated with the
corresponding amplitude probability $\phi(\vec{p})$ to find a quark with
momentum $\vec{p}$ inside a meson. In the center-of-mass reference frame,
the $S$ matrix at second order takes the form:

\begin{eqnarray}
S & = & -i e^2 e_q (2 \pi)^4 \int \frac{\delta^{(4)}(P_i - P_f) m_l m_q} {(2
\pi)^3 E_p (2 \pi)^3 E_q}  \nonumber \\
& & \sum_{M_1 M_2 \mu_L \mu_S} \langle L \mu_L S \mu_S \vert J \mu_J \rangle
\langle \frac{1}{2} M_1 \frac{1}{2} M_2 \vert S \mu_S \rangle \, d^3 p \,
\phi(\vec{p}) \, \frac{g_{\mu \nu}}{M^2} A_l^{\mu} A_q^{\nu},
\label{inicial}
\end{eqnarray}
where $A_l^{\mu} \equiv [ \bar{u}_l(q,\xi_1) \gamma^\mu v_l(-q,\xi_2) ]$
denotes the leptonic part of the $S$ matrix and $A_q^{\nu} \equiv [ \bar{v}%
_q(-p,M_2) \gamma^\nu u_q(p,M_1) ]$ the quark part. In this expression we
used the fact that the relative momentum $\vec{p}$ appearing in the meson
wave function is chosen to be the same as the quark momentum and opposite to
the antiquark momentum.

Let us first compute the quark term. Taking for the Dirac spinors the
expressions

\begin{eqnarray}
v_q^\dag & = & \sqrt{\frac{E_p+m_q}{2 m_q}} \left( -\Phi^\dag_{-M_2}\frac{%
\vec\sigma \cdot \vec p}{E_p+m_q} , \Phi^\dag_{-M_2} \right),  \nonumber \\
u_q & = & \sqrt{\frac{E_p+m_q}{2 m_q}} \left( 
\begin{array}{c}
\Phi_{M_1} \\ 
\frac{\vec\sigma \cdot\vec p}{E_p+m_q} \Phi_{M_1}
\end{array}
\right),  \label{inicial2}
\end{eqnarray}
we obtain

\begin{equation}
A_q^\nu = \frac{E_p+m_q}{2 m_q} \left( -\Phi^\dag_{-M_2}\frac{\vec\sigma
\cdot\vec p} {E_p+m_q} , \Phi^\dag_{-M_2} \right) \gamma^0 \gamma^\nu \left( 
\begin{array}{c}
\Phi_{M_1} \\ 
\frac{\vec\sigma \cdot\vec p}{E_p+m_q} \Phi_{M_1}
\end{array}
\right).
\end{equation}

In the case $\nu=0$ the result is identically zero. For technical
simplicity, we switch from cartesian to spherical coordinates, so from now
on the latin indices will run to $-$1, 0, 1. Using the relation $(\vec\sigma
\cdot\vec p) \sigma_i (\vec\sigma \cdot\vec p)= 2 p_i (\vec\sigma \cdot\vec p%
) - p^2 \sigma_i$ the quark part of the $S$ matrix can be written as,

\begin{equation}
A_q^i = \frac{E_p+m_q}{2m_q} \left[\left( 1 + \frac{p^2}{(E_p+m_q)^2}
\right) \Phi^\dag_{-M_2} \sigma_i \Phi_{M_1} -2 \sum_j \frac{p_i p_j^\ast}{%
(E_p+m_q)^2} \Phi^\dag_{-M_2} \sigma_j \Phi_{M_1} \right].  \label{eq-5}
\end{equation}

As seen in this equation, to evaluate the $S$ matrix the expectation value
of the Pauli matrices between spin eigenstates ($\Phi_M$) is needed. These
functions are not eigenstates of the Dirac free hamiltonian. Moreover, they
are eigenfunctions of $\sigma_0$ only for small values of the momentum $p$
(but the contributions of large momenta are suppressed by the exponential
tail of the wave function). Nevertheless, the calculation of $A_q^i$ can
still be done with the $\Phi_M$ spinors, but being very careful with phases.
Alternatively, one can switch to the helicity scheme, for which the spinors
are eigenstates of the Dirac free hamiltonian. We will work in the last
scheme, but we checked that both methods give exactly the same results.

Given a spin one-half particle with momentum $\vec{p}= (p, \theta, \varphi)$%
, the relation between spin and helicity eigenstates is \cite{varsha}:

\begin{eqnarray}
\vert \Phi_{M} \rangle & = & \sum_{\lambda} D^{1/2}_{-\lambda, -M} (0,
\theta, \varphi) \vert \chi_\lambda (\theta, \varphi) \rangle,  \nonumber \\
\langle \Phi_{M} \vert & = & \sum_{\lambda} (-1)^{\lambda-M}
D^{1/2}_{\lambda, M} (0, \theta, \varphi) \langle \chi_\lambda (\theta,
\varphi) \vert ,
\end{eqnarray}
where $D^{1/2}_{m_1,m_2} (0, \theta, \varphi)$ are the Wigner matrix
elements. For a two-particle state with momenta $\vec{p}$ and $-\vec{p}$, in
order to keep the system invariant under Lorentz transformations, it is
necessary to introduce an additional phase in the wave function of the
second particle \cite{jacob}. This phase is $(-1)^{j-\lambda}$, being $j$
the particle spin and $\lambda$ its helicity, and therefore the spinors take
the form

\begin{eqnarray}
\vert \Phi_{M_1} \rangle & = & \sum_{\lambda_1} D^{1/2}_{-\lambda_1, -M_1}
(0, \theta, \varphi) \vert \chi_{\lambda_1} (\theta, \varphi) \rangle, 
\nonumber \\
\langle \Phi_{-M_2} \vert & = & \sum_{\lambda_2} (-1)^{1/2+M_2}
D^{1/2}_{\lambda_2, -M_2} (0, \theta, \varphi) \langle \chi_{\lambda_2}
(\theta, \varphi) \vert .
\end{eqnarray}

Using the Wigner matrix elements and the mean value of the Pauli matrices
between helicity states given in Table \ref{ta-pauli}, and summing over all
helicity states, we get the following results for $\Phi^\dag_{-M_2} \sigma_i
\Phi_{M_1}$ as a function of the spin indices $M_1$ and $M_2$,

\begin{center}
\begin{tabular}{c|cccc}
$(M_1,M_2)$ & $(1/2,1/2)$ & $(1/2,-1/2)$ & $(-1/2,1/2)$ & $(-1/2,-1/2)$ \\ 
\hline
$\Phi^\dag_{-M_2} \sigma_i \Phi_{M_1}$ & $-\sqrt{2} \delta_{i,-1}$ & $%
\delta_{i,0}$ & $\delta_{i,0}$ & $-\sqrt{2}\delta_{i,1}$.
\end{tabular}
\end{center}

One can then perform the summation over $M_1$ and $M_2$ in Eq. (\ref{inicial}%
) arriving to

\begin{equation}
\sum_{M_1 M_2} \langle \frac{1}{2} M_1 \frac{1}{2} M_2 \vert S \mu_S \rangle
\Phi^\dag_{-M_2} \sigma_i \Phi_{M_1} = \sqrt{2} (-1)^{i+1} \delta_{S,1}
\delta_{i,-\mu_S},
\end{equation}

\noindent from where a first selection rule for the leptonic decay: $S=1$,
is obtained. Eq. (\ref{eq-5}) can be finally written as

\begin{equation}
A_q^i = \delta_{S,1} \sqrt{2} \left[\frac{E_p}{m_q} (-1)^{i+1}
\delta_{i,-\mu_S} - \frac{1}{m_q(E_p+m_q)}\sum_j p_i p_j^\ast (-1)^{j+1}
\delta_{j,-\mu_S} \right],
\end{equation}
where it has been used that

\begin{equation}
1+\frac{p^2}{(E_p+m_q)^2} = \frac{2 E_p}{E_p+m_q}.
\end{equation}

Let us now concentrate on the part of the $S$ matrix which depends on the
quark and the antiquark,

\begin{equation}
B^i \equiv \int d^3p \phi(\vec{p}) \sum_{\mu_L \mu_S} \langle L \mu_L S
\mu_s \vert J \mu_J \rangle \delta_{S,1} \sqrt{2} \left[(-1)^{i+1}
\delta_{i,-\mu_S} - \frac{1}{E_p(E_p+m_q)}\sum_j p_i p_j^\ast (-1)^{j+1}
\delta_{j,-\mu_S} \right].  \label{medio}
\end{equation}

The integration of the first term can be done through the relation $\int
d^3p \phi(\vec{p}) = (2 \pi)^{3/2} \Psi(0)$, arriving to

\begin{equation}
(2 \pi)^{3/2} \Psi(0) \delta_{L,0} \delta_{S,1} \delta_{J,1} \delta_{i,
-\mu_J} \sqrt{2} (-1)^{1-\mu_J}.
\end{equation}

The integration of the second term is a little bit more involved, and after
some algebra one arrives to

\begin{equation}
- \frac{\sqrt{2} \sqrt{2} \sqrt{2L+1} (2\pi)^{3/2}}{6 \pi} \langle L 0 1 0
\vert 1 0 \rangle I_4 \delta_{S,1} \delta_{J,1}
\delta_{i,-\mu_J}(-1)^{1-\mu_J},
\end{equation}
where we have defined

\begin{equation}
I_4 \equiv \int \frac{p^4 dp R_p(p)}{E_p(E_p+m_q)}.
\end{equation}

Taking the two parts together, Eq. (\ref{medio}) is finally written as

\begin{equation}
B^i = (2\pi)^{3/2} \delta_{S,1} \delta_{J,1} \delta_{i, -\mu_J} \sqrt{2}
(-1)^{1-\mu_J} \left[ \Psi(0) \delta_{L,0} - \frac{\sqrt{2}}{6 \pi} \sqrt{%
2L+1} \langle L 0 1 0 \vert 1 0 \rangle I_4 \right].  \nonumber
\end{equation}

The $\delta$ function and the Clebsch-Gordan coefficient provide a new
selection rule: $L=0,2$. The first term between brackets gives rise to the
classical van Royen-Weisskopf formula, it only concerns to $^3S_1$
resonances. The second term is the modification due to non-zero quark
momentum distribution inside the meson (through the integration over $p$ in $%
I_4$) and it also contains kinematical relativistic effects (through the $%
E_p $ denominator in $I_4$). It contributes to $L=0$ ($^3S_1$) resonances
and $L=2$ ($^3D_1$) resonances, whose decays are strictly forbidden within
the original van Royen-Weisskopf formalism. The recipe for taking into
account the previous improvements is quite simple; one makes the substitution

\begin{equation}  \label{modif}
\Psi(0) \rightarrow \Psi(0) \delta_{L,0} - \frac{\sqrt{2}}{6 \pi} \sqrt{2L+1}
\langle L 0 1 0 \vert 1 0 \rangle I_4 \equiv \overline{\Psi(0)}
\end{equation}

\noindent in the traditional van Royen-Weisskopf formula.

Finally, we have for the $S$ matrix,

\begin{equation}  \label{dsif}
S = i e^2 e_q \frac{1}{(2 \pi)^2} \frac{m_l}{E_q} \frac{1}{M^2} \sqrt{2}
(2\pi)^{3/2} \overline{\Psi(0)} \delta^{(4)}(q_1+q_2-P) \sum_i A_l^{-i}
\delta_{i,-\mu_J}.
\end{equation}

The rest of the calculation to obtain the expression of the width is
detailed in appendix \ref{ap-lep}. Up to now, the $S$ matrix has been
calculated for a quark-antiquark system with given colour and flavour.
Considering that the meson is a colour singlet with a contribution of each
colour with the same probability, the net colour factor is a multiplication
by 3 as compared to the usual van Royen-Weisskopf expression. For neutral
ordinary mesons composed of two flavours $f$ ($u$ and $d$) with quark charge 
$e_q(f)$, the flavour wave function is written generally as $\sum_{f}a_f f {%
\bar f}$ (the amplitude $a_f$ on the $f {\bar f}$ channel is simply a
Clebsch-Gordan coefficient). Provided an SU(2) invariance of the potential,
the meson wave function is the same for the two flavour channels, so that
flavour degree of freedom implies only the replacement $e_q \rightarrow
\sum_{f} a_f e_q(f)$. The final result for the width is:

\begin{equation}
\Gamma = \frac{16 \pi \alpha^2}{M^2} \left( 1 - 4 \frac{m_l^2}{M^2}
\right)^{1/2} \left(1 + 2 \frac{m_l^2}{M^2} \right) (\sum_{f} a_f e_q(f))^2
\vert \overline{\Psi(0)} \vert^2,  \label{eq-gam}
\end{equation}
being $\alpha$ the fine structure constant.

\subsection{Electromagnetic size for quarks}

\label{sec-ele}

In the previous section we have treated quarks as point-like charge
distributions. However, the constituent quark is a complicated object
surrounded by a quark-antiquark pair cloud. Vogl et al.\cite{vogl} have
explicitly shown in a generalized Nambu-Jona-Lasinio  model that constituent quarks are
dressed by a mesonic ($q\bar{q}$) polarization cloud. For a given external
field the $q\bar{q}$ modes,
 which carry the quantum numbers of the probe, will
determine the size of the constituent quark. Hence, the
electromagnetic properties of constituent quarks in the low-energy regime 
are largely determined by the screening effects induced by $q\bar{q}$
polarization cloud with $\rho $ meson quantum numbers, 
and the old vector-dominance
principle emerges naturally.

Based on these results several authors have studied hadronic
electromagnetic properties assuming that the electromagnetic size of the
constituent quark can be described by a monopole form factor 
\begin{equation}
F(q^{2})=\frac{1}{1+q^{2}m^{2}}.
\end{equation}
Magnetic form factors \cite{buch1}, magnetic moments \cite{wag},
electromagnetic properties of the $\Delta (1232)$\cite{buch2}, pion form
factors \cite{card1} and radiative $\pi \rho $ form factors \cite{card2},
among others, have been calculated in different models. The results show the
relevance of the quark electromagnetic size.

The quark electromagnetic size is then related to the corresponding
vector-meson propagator and it has a dependence on the inverse of the
squared vector-meson mass. In this way, the quark electromagnetic size is
flavour-dependent, through the flavour dependence of the vector-meson mass.
Instead of using the preceding expresion,
we simulate the quark electromagnetic size by a
density $\rho _{q}(\vec{r})$. These quarks will be called from now on {\it %
dressed quarks}. Let us denote by $a^{\dag }(\vec{r})$ the creation operator
of a bare quark at position $\vec{r}$; in the same way $A^{\dag }(\vec{r})$
stands for the creation operator of a dressed quark at position $\vec{r}$.
It seems natural to define:

\begin{equation}
A^\dag (\vec{r}) = \int d^3 r^{\prime}\rho_q (\vec{r} - \vec{r}\ ^{\prime})
a^\dag (\vec{r}\ ^{\prime}).
\end{equation}

If the density function $\rho_q(\vec{r}) = \delta^{(3)}(\vec{r})$ , we have $%
A^\dag(\vec{r}) = a^\dag(\vec{r}) $ and we recover bare quarks. Strictly
speaking, the above equation is incorrect and misleading, in particular
commutation relations among $A^\dag (\vec{r})$ operators are no longer
satisfied. Nevertheless, we still apply this equality keeping in mind that
it makes sense only when the $\rho_q (\vec{r} - \vec{r}\ ^{\prime})$
distribution is a sharply peaked function. This is the price to pay to get a
phenomenological very simple modification to a finite size for the quarks.

Let us calculate now the Fourier transform of the densities and of the
creation operators. Defining

\begin{eqnarray}
\tilde{A}^\dag (\vec{p}) & = & \frac{1}{(2\pi)^{3/2}} \int d^3 r A^\dag(\vec{%
r}) e^{i\vec{p}.\vec{r}}  \nonumber \\
\tilde{a}^\dag (\vec{p}) & = & \frac{1}{(2\pi)^{3/2}} \int d^3 r a^\dag(\vec{%
r}) e^{i\vec{p}.\vec{r}}  \nonumber \\
\tilde{\rho_q} (\vec{p}) & = & \frac{1}{(2\pi)^{3/2}} \int d^3 r \rho_q(\vec{%
r}) e^{i\vec{p}.\vec{r}},
\end{eqnarray}

\noindent from the properties of the Fourier transformations it is immediate
to see that:

\begin{equation}
\tilde{A}^\dag(\vec{p}) = (2 \pi)^{3/2} \tilde{\rho_q}(\vec{p}) \tilde{a}%
^\dag(\vec{p}).
\end{equation}

We define (forgetting the obvious complications due to other degrees of
freedom and the corresponding couplings) the creation operators of a meson
as a bare or dressed quark-antiquark pair creation, respectively:

\begin{eqnarray}
d^\dag & = & \int d^3 p \, \phi(\vec{p}) \, \tilde{a}^\dag (\vec{p}) \tilde{b%
}^\dag (-\vec{p}),  \nonumber \\
D^\dag & = & \int d^3 p \, \phi(\vec{p}) \, \tilde{A}^\dag (\vec{p}) \tilde{B%
}^\dag (-\vec{p}) ,
\end{eqnarray}
where $b^\dag$ and $B^\dag$ are the corresponding creation operators for
bare and dressed antiquarks, respectively.

Operating, we get

\begin{eqnarray}
D^\dag & = & (2 \pi)^3 \int d^3p \, \phi(\vec{p}) \, \tilde{\rho_q}(\vec{p}) 
\tilde{\rho_{\bar{q}}}(-\vec{p}) \tilde{a}^\dag(\vec{p}) \tilde{b}^\dag(-%
\vec{p})  \nonumber \\
& = & (2 \pi)^3 \int d^3p \, \phi(\vec{p}) \, \tilde{\rho_{q\bar{q}}}(\vec{p}%
) \tilde{a}^\dag(\vec{p}) \tilde{b}^\dag(-\vec{p}) ,
\end{eqnarray}
where we have defined

\begin{equation}
\tilde{\rho_{q\bar{q}}}(\vec{p}) = \tilde{\rho_q}(\vec{p}) \tilde{\rho_{\bar{%
q}}}(-\vec{p}).
\end{equation}

Note that, since the antiquark has the same mass as the quark and the
densities depend only on $p^2$, it is natural to take $\tilde{\rho_q}(\vec{p}%
)=\tilde{\rho_{\bar{q}}}(-\vec{p})$, so that:

\begin{equation}
\tilde{\rho_{q\bar{q}}}(p) = \tilde{\rho_q}(p)^2,
\end{equation}
which implies 
\begin{equation}
\tilde{\rho_{q\bar{q}}}(\vec{p}) = \tilde{\rho_{q\bar{q}}}(-\vec{p}).
\end{equation}

When operating on the vacuum state $\vert 0 \rangle$, $D^{\dag}$ gives the
spinors $u_q $ and $\bar{v}_q$ of Eq. (\ref{inicial2}). In the previous
section, we arrive to Eq. (\ref{medio}), in which there are two terms. The
only dependence of the first term in $p$ is in the wave function; with
dressed quarks we would have

\begin{eqnarray}
(2 \pi)^3 \int d^3p \, \phi(\vec{p}) \, \tilde{\rho_{q\bar{q}}}(\vec{p}) & =
& (2 \pi)^3 \int d^3p \, \phi(\vec{p}) \, \tilde{\rho_{q\bar{q}}}(-\vec{p}) 
\nonumber \\
& = & (2 \pi)^3 \int d^3p \, \phi(\vec{p}) \, \frac{1}{(2 \pi)^{3/2}} \int
d^3r e^{-i\vec{p}.\vec{r}} \rho_{q\bar{q}}(\vec{r})  \nonumber \\
& = & (2 \pi)^3 \int d^3r \, \Psi(\vec{r}) \, \rho_{q\bar{q}}(\vec{r}),
\end{eqnarray}
while the second term would simply include the quark density.

Therefore, when introducing dressed quarks, it is necessary to substitute:

\begin{eqnarray}
\Psi(0) & \rightarrow & (2 \pi)^{3/2} \int d^3r \rho_{q\bar{q}} (\vec{r})
\Psi(\vec{r}), \\
I_4 \equiv \int \frac{p^4 dp R_p(p)}{E_p(E_p+m_q)} & \rightarrow & (2 \pi)^3
\int \frac{p^4 dp R_p(p)}{E_p(E_p+m_q)} \tilde{\rho_{q\bar{q}}}(p).
\end{eqnarray}

\section{Meson wave functions}

\label{sec-mod} Once the mechanism to calculate the decay widths has been
described, the only remaining point is the meson wave function. The
sensitivity of the results on the meson wave function will be analyzed by
comparing two different potential models which seem to us very
representative of the high degree of sophistication reached in meson
spectroscopy. The first one is the chiral constituent quark model, $\chi$QM.
The central feature of the model is the concept of constituent quark mass,
which appears as a consequence of spontaneous chiral symmetry breaking. The
potential, derived and discussed elsewhere \cite{bo-bl,jpg}, includes a
confining interaction, a perturbative one-gluon exchange and Goldstone-boson
exchanges. It can be easily extended to the study of mesons in which at
least one of the components is a $c$ or $b$ quark (or antiquark). In this
case, chiral symmetry is explicitly broken and the potential only includes
confinement and one-gluon exchange. The parameters used are the same as in
Ref. \cite{bo-bl}, with the addition of the charm and bottom masses: 1845
and 5250 MeV, respectively. This potential model provides reasonable results
for the baryon-baryon interaction \cite{jpg}, baryon spectra \cite{bs}, meson
spectra \cite{ms} and meson strong decays \cite{bo-bl}. As a consequence,
the calculation of meson leptonic decays constitutes a severe test of the
model as it supposes a parameter-free calculation.

The second potential, called DNR, has been developed around gluon exchanges 
\cite{sem} and relies very closely to the fundamental QCD. The short-range
part sticks to the experimental $\alpha_s(q^2)$ coupling constant while the
long-range confinement is essentially linear with a string tension roughly
equal to the experimental one. Moreover, it includes instanton induced
effects, which play a crucial role in explaining the isoscalar spectra, and
hyperfine terms as a sum of two gaussian functions. The quarks have also
been dressed with a gluonic cloud (which has nothing to do with the
electromagnetic cloud considered above), so that the bare potential is
convoluted with a gluonic density to give the final potential. The free
parameters have been fitted in all mesonic sectors and the corresponding
spectra are quite good.

\begin{figure}[tb]
\begin{center}
\includegraphics{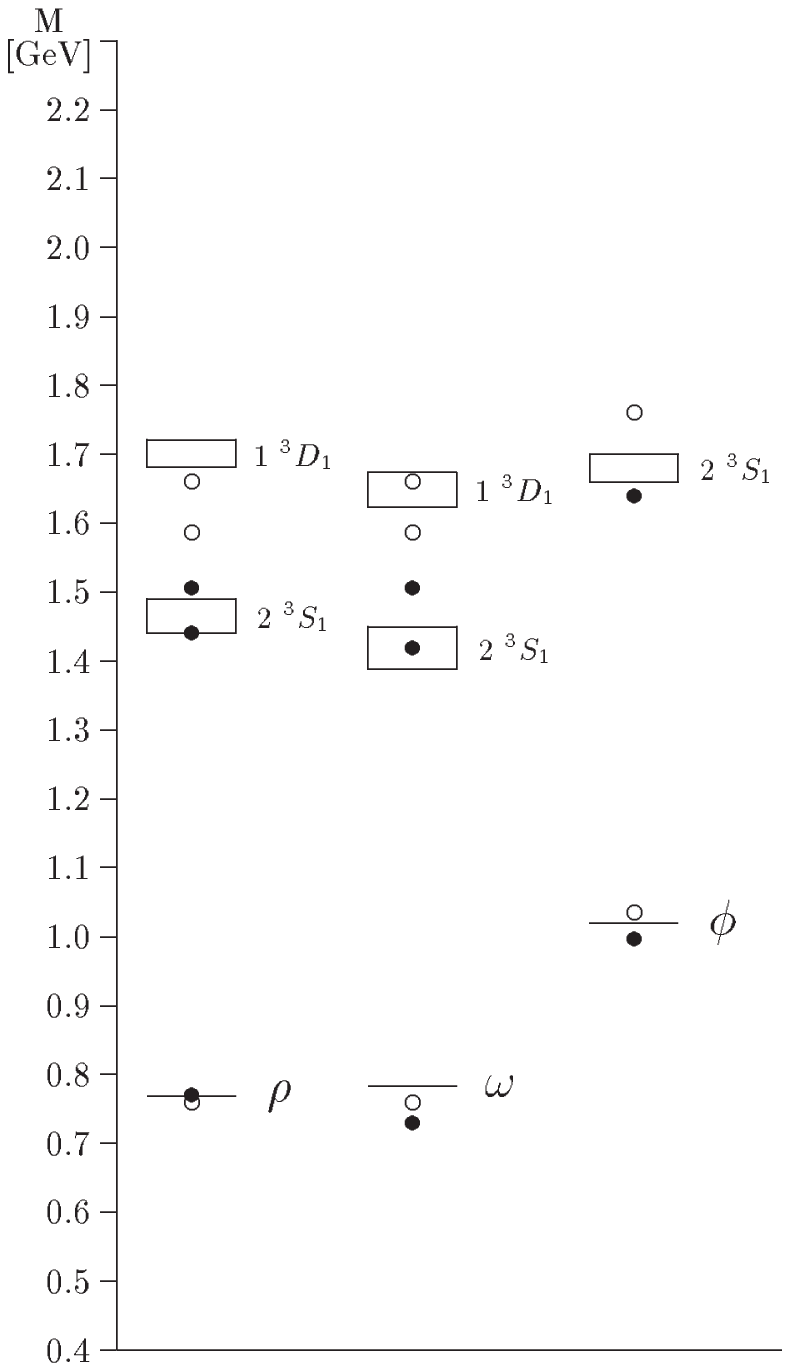}
\includegraphics{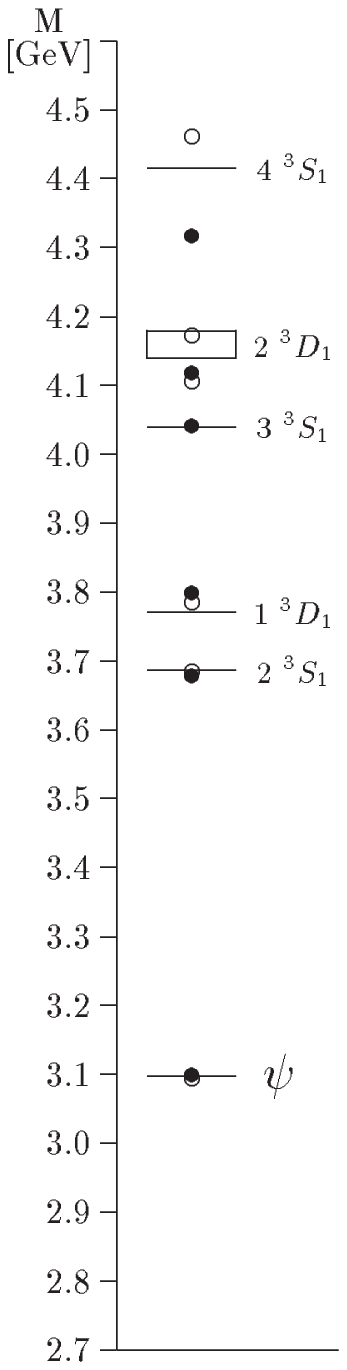}
\includegraphics{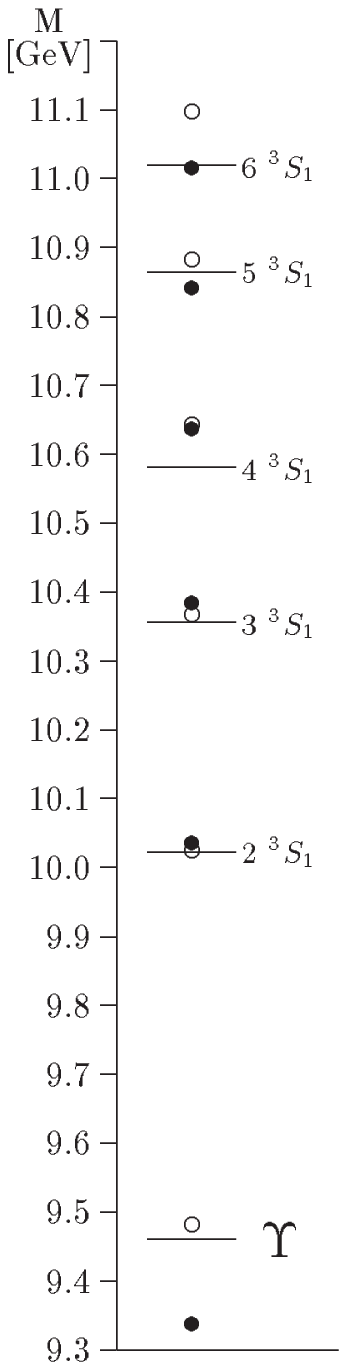}
\end{center}
\caption{Comparison of the DNR and $\protect\chi$QM spectra for $J^P=1^-$
resonances discussed in the text. The horizontal lines indicate the
experimental values of the masses, being rectangles in the case of an
experimental uncertainty bigger than 20 MeV, the open circles show the
values obtained with DNR potential and the dark circles those from $\protect%
\chi$QM potential. The $\protect\rho$($n\bar{n}$ system, I=1), $\protect%
\omega$($n\bar{n}$ system, I=0), $\protect\phi$($s\bar{s}$ system) families
have a common scale whereas $\protect\psi$($c\bar{c}$ system) and $\Upsilon$(%
$b\bar{b} $ system) families have their own one. The spectroscopic quantum
numbers are indicated close to the experimental mass values. }
\label{spectre}
\end{figure}

Both models rely on a non-relativistic expression of the kinetic energy
operator and they need to solve the Schr\"odinger equation (for example
using a Numerov algorithm). To compare their respective merits we have
plotted in Fig. \ref{spectre} the spectra for the resonances appearing in
this study. The quality is more or less the same, some resonances being
better reproduced with $\chi$QM and others with DNR. In particular, except
for the $\Upsilon$, which raises a problem to $\chi$QM, both potentials are
able to reproduce the heavy sector (charmonium and bottomonium) rather
nicely, even for very excited levels (up to $6\ ^3S_1$ in bottomonium). In
the light sector (ordinary and strange) the situation is a little less
favorable; in general $\chi$QM reproduces better first radial excited
states, while DNR looks better for orbital excitations.

\begin{figure}[tb]
\begin{center}
\mbox{
\includegraphics[width=0.8\linewidth]{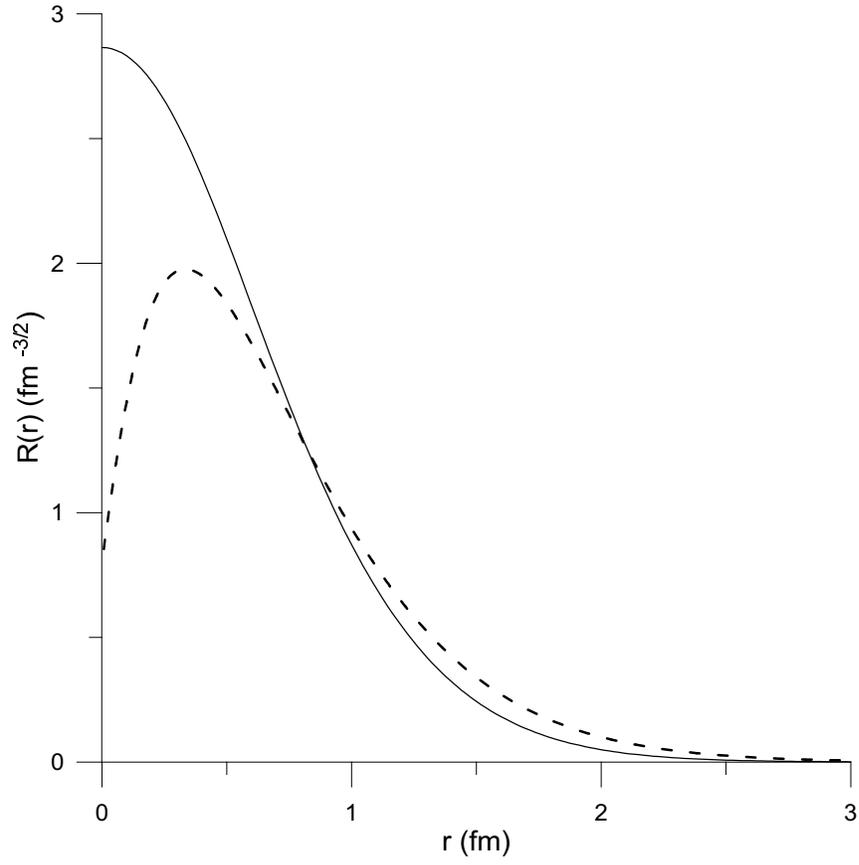}}
\end{center}
\caption{Radial wave functions $R(r)$ of the $\protect\rho$ meson obtained
with DNR (solid line) and $\protect\chi$QM (dashed line) potentials. }
\label{rho}
\end{figure}

Although both potentials give spectra of roughly comparable quality, they
can provide rather different wave functions. As an illustration of this
fact, we draw in Fig. \ref{rho} the wave function $R(r)$ of the $\rho$
resonance. $\chi$QM wave function shows a hole near the origin (due to a
stronger hyperfine term) which is absent in DNR wave function. As a
consequence there is a factor $\approx 3$ difference for the value of the
wave function at the origin (and thus a factor around 10 for the width). We
thus see the fundamental importance to rely on observables other than
energies to test the quality of a model.

\section{Results and discussion}

\label{sec-res}

The previous formalism is applied to all experimentally known decays of a
meson into a lepton-antilepton pair. Some other transitions have been seen
experimentally but without precise data; we present also our results in this
case as a matter of prediction.

In order to isolate the dynamical effects from the kinematical ones, we
employ the experimental meson masses for the phase space factor. Let us
emphasize that the calculation of the meson leptonic decay widths in the
bare quark case is parameter-free, all the model parameters being fixed to
other observables (meson spectra for DNR, and $NN$ interaction and meson
spectra for $\chi$QM). The results dealing with the dressed quarks need to
determine the size $\sigma$ of the electromagnetic distribution. As detailed
in appendix \ref{ap-gau}, we use a gaussian charge density and the flavour
dependence of the size is parametrized by

\begin{equation}
\sigma_f = \sigma_0 e^{m_f/m_0},
\end{equation}
where $m_f$ is the constituent quark mass of the corresponding flavour $f$,
and the parameters $\sigma_0$ and $m_0$ are fitted on the known transitions.
The obtained values are $(\sigma_0,m_0)$ = (0.6 GeV, 1.25 GeV) and (0.7 GeV,
2.11 GeV) for DNR and $\chi$QM, respectively. It is worth to notice that the
electromagnetic sizes coming from this fit are in good agreement with the
prediction of vector-meson dominance model.

The results for leptonic decay widths are presented in Table \ref{ta-re}. We
compare the value obtained for the $\chi$QM and DNR meson wave functions in
three different cases. The first two columns report the results derived from
the original van Royen-Weisskopf formula. Columns three and four show the
results when the modified van Royen-Weisskopf formula of Eq. (\ref{modif})
is used. Finally, columns five and six contain the results when, on top of
the modified van Royen-Weisskopf formula, the dressing for the quarks is
used.

Let us comment first the values due to a pure van Royen-Weisskopf formula
(VR). In this case, we discard the decay of $^3D_1$ resonances that is
strictly forbidden. The DNR potential systematically gives a too large wave
function at the origin; as a consequence the theoretical width can exceed
the experimental one by a factor 3, but in the bottomonium sector the values
are much closer to the experimental data. Moreover, the hierarchy in the
relative importance is respected, indicating VR contains already a part of
truth. More puzzling is the case of $\chi$QM potential. An important
difference can be established between the decay results for light mesons ($%
\rho$, $\omega$ and $\phi$) which are smaller than the experimental data and
heavy mesons ($\psi$ and $\Upsilon$) which are in general larger than the
experimental ones. This is obviously due to the hole in the wave function
near the origin (see Fig. \ref{rho}). Although all of them have $S=1$, and
therefore the spin-spin contact term included in the one-gluon exchange is
repulsive, for light mesons this term is strong, as it depends on the
inverse product of quark masses, and makes the wave function to have its
maximum at $r \approx $ 0.5 fm. As a consequence, the van Royen-Weisskopf
value for these decay widths is smaller than experiment. On the other hand,
for heavy mesons this repulsion is much less important, and the wave
function takes its maximum at (nearly) the origin. The van Royen-Weisskopf
value for decays of these mesons is higher than experimental ones, similarly
to the results obtained previously in the literature. For the heavy sector $%
\chi$QM gives poorer results than DNR.

For heavy quarkonia, in both models, our extension of the formalism to
include non-zero momentum for the quarks and kinematical relativistic
effects gives always a negative contribution. The difference with
non-relativistic results is between 10 and 30 \%, bigger than in Ref. \cite
{bey} but much smaller than in Ref. \cite{lou}. The main effect is thus to
decrease the theoretical value, making it closer to experiment. The same
conclusion applies for light mesons with DNR wave functions. The case of
light mesons with $\chi$QM is more interesting. The modification is
positive, so the transition is enhanced making it closer to experiment
again. Thus, whatever the importance of the wave function at the origin is,
the modification presented above always improves the theoretical results.
This indicates again that there is certainly a part of truth in this
explanation. In addition, we get non-zero widths for $D$-wave mesons, like $%
\psi(3770)$ and $\psi(4160)$, although the calculated widths are much
smaller than the experimental ones, as also observed in Ref. \cite{bey}.

The most significative improvement is produced when the electromagnetic size
of quarks is considered. As the term included in the van Royen-Weisskopf
formula is the most relevant to obtain the width, the quark finite size
replaces the value of the wave function at the origin (which is strongly
dependent on the structure of the potential at very small distances) by an
average on the whole wave function, taking into account the behaviour of the
system everywhere. The $I_4$ integral is also sensitively changed. For DNR
the main effect is to still reduce the theoretical values and make them very
close to the experimental data. The results in this case can be considered
as very good, owing to the fact that we have only two free parameters at our
disposal. For $\chi$QM the behaviour depends again on which sector we
consider. For the heavy mesons, the dressing of the quarks reduces the
theoretical value and makes it closer to experiment. For light mesons the
width is importantly increased, being also close to the experimental value.
Thus, in any case, the description is improved. Although the results of $%
\chi $QM are slightly worse than those of DNR, the conclusion is that the
dressing is very important and it always improves the situation.

The dressing does not affect very much the decay of $D$-wave resonances,
whose calculated widths remain still too low. We think that this point must
deserve further considerations like the inclusion of tensor forces which are
outside the scope of this work.

Concerning the branching ratios between the different lepton channels ($e^+
e^-$, $\mu^+ \mu^-$ and $\tau^+ \tau^-$) our formalism always gives bigger
widths for lighter leptons, while the experimental results are in almost all
cases the opposite. For the moment, we have no explanation concerning this
point.

One remark should be done about the ratio between $\rho $ and $\omega $
widths. If both mesons were degenerated and had the same radial wave
function, this ratio would take the value of 9 due to flavour content. Since 
$\rho $ is experimentally lighter than $\omega $, there exists already a
small deviation due to phase space factor. One would expect also that its
wave function would be slightly higher at the origin and the ratio would be
closer to the experimental value, which is approximately 11. In DNR the wave
functions are absolutely identical, so that this dynamical effect does not
exist. With $\chi $QM the situation is different because the potential is
isospin dependent; nevertheless, we find $m_{\rho }>m_{\omega }$ and
therefore $\Psi (0)_{\rho }<\Psi (0)_{\omega }$; as a result, the calculated
value for the ratio is even smaller than 9.

With respect to the transitions for which precise data are not known, both
models give results which are relatively close to each other, and which, in
any case, respect the hierarchy of importance. We can thus have some
confidence on them.

\begin{figure}[tb]
\begin{center}
\mbox{
\includegraphics[width=0.8\linewidth]{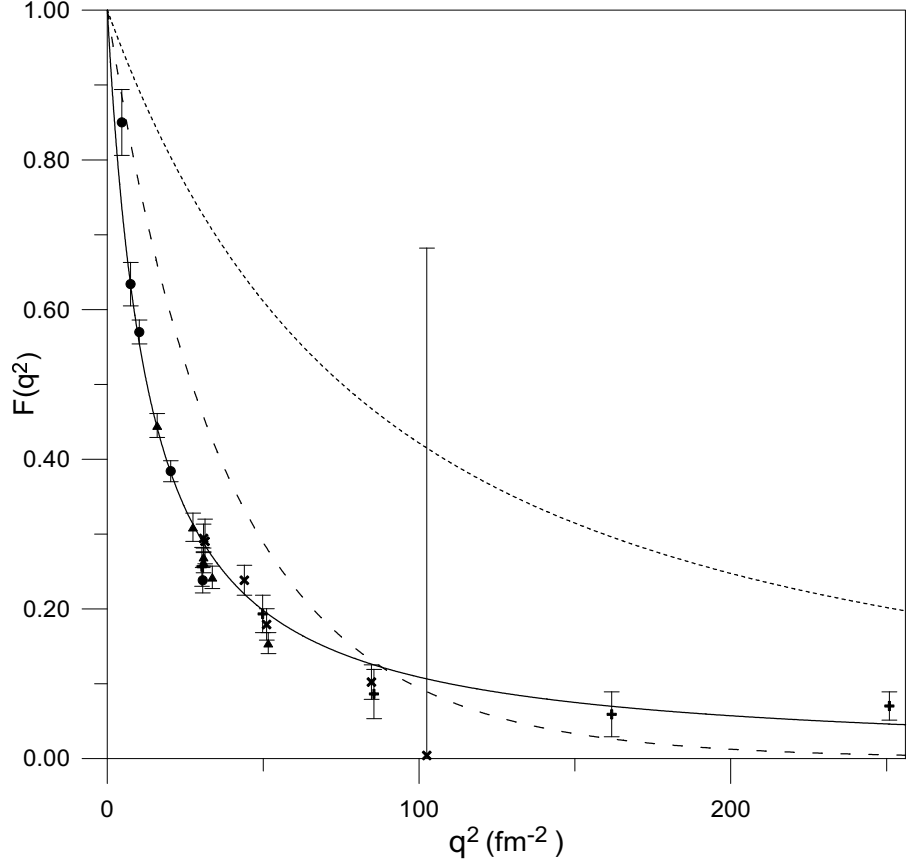}}
\end{center}
\caption{Pion charge form factor
calculated with the $\chi$QM wave function. 
The dotted line corresponds to the
bare quark calculation. Dashed line corresponds to the dressed
quark calculation. Solid line includes the effect of boosting.}
\label{fig-ff}
\end{figure}

Finally, we would like to show how the quark electromagnetic size
we have used  works in
other electromagnetic observables. In Fig. \ref{fig-ff} we present the pion charge form
factor calculated with the $\chi$QM wave function used in this work. 
The dotted line
represents the calculation done with bare quarks whereas
the dashed line corresponds to the
dressed quarks calculation. As can be seen, the inclusion of the quark
electromagnetic size improves the agreement with the experimental data.
Although it is not our aim to make any definitive statement about this
problem we can introduce the effect of boosting on the form factor (which has
been calculated using rest frame wave functions) using the prescription of
Stanley and Robson \cite{rob} to take into account the kinematical meson
recoil. The result of this boosting is shown by the solid line,
obtaining a
reasonable description of the data. It is worth to notice that the
calculation has been done in the one-body approximation framework for the
electromagnetic current, without taking into account the two-body currents
necessary for fulfilling gauge invariance. However, it is clear that a
reasonable agreement with the data can be achieved by assuming an effective
one body current between constituent quarks with non-trivial
electromagnetic structure. This conclusion agrees with the calculation of 
Ref. \cite{card1}, altough in a different framework.

\section{Summary}

In this paper we investigate whether the van Royen-Weisskopf formula for
meson leptonic decays can be improved. We first include the possibility of
non-zero momentum for the quark-antiquark pair within the meson as well as
kinematical relativistic effects. We also introduce phenomenologically an
electromagnetic density for dressing the quarks, that we choose of gaussian
form. We make our study on all experimentally known transitions and use wave
functions coming from two different non-relativistic quark model potentials
describing correctly the meson spectra. The first one, $\chi$QM, is based on
gluon and Goldstone-boson exchanges, while DNR contains gluon exchange as
well as instanton induced effects. Although the wave functions coming from
both models can appreciably differ, we showed that the proposed
modifications always go in the right direction.

The inclusion of non-zero momentum for quarks inside the meson gives an
effect of order 10\%-30\%, while the inclusion of a quark 
electromagnetic size improves a
lot the situation. With two adjustable parameters, our theory is able to
reproduce nicely the experimental data for transitions coming from $S$-wave
resonances. Our proposed modifications give non-zero width for $D$-wave
mesons, but seem to be not enough to obtain their correct order of
magnitude. The fact that they systematically improve the description shows
that we certainly introduce basically good ingredients.

\section{acknowledgments}

\indent
This work has been partially funded by Direcci\'on General de
Investigaci\'on Cient\'{\i}fica y T\'ecnica (DGICYT) under the Contract No.
PB97-1410, by Junta de Castilla y Le\'on under the Contract No. SA109/01 and
by a IN2P3-CICYT agreement.
A.V. thanks the Ministerio de Educaci\'on, Cultura y Deporte of Spain for
financial support through the Salvador de Madariaga program.

\appendix

\section{Leptonic part and final calculation of the decay width}

\label{ap-lep}

We start from Eq. (\ref{dsif}), making explicit the leptonic part 
\begin{eqnarray}
S & = & i e^2 e_q \frac{1}{(2 \pi)^2} \frac{m_l}{E_q} \frac{1}{M^2} \sqrt{2}
(2\pi)^{3/2} \overline{\Psi(0)} \delta^{(4)}(q_1+q_2-P) \sum_i
\delta_{i,-\mu_J}  \nonumber \\
& & \frac{E_q+m_l}{2 m_l} \left[ \left(1 + \frac{q^2}{(E_q+m_l)^2} \right)
\Phi^\dag_{\xi_1} \sigma_{-i} \Phi_{-\xi_2} - 2 \sum_k (-1)^k
\Phi^\dag_{\xi_1} \sigma_{k} \Phi_{-\xi_2} \frac{q_{-i}q_{-k}}{(E_q+m_l)^2} %
\right].
\end{eqnarray}

For the leptonic part, as we are performing a sum over all the final states,
the same result is obtained when using spin or helicity eigenstates. We take
spin states for simplicity. Using the Wigner-Eckart theorem,

\begin{equation}
\Phi^\dag_{\xi_1} \sigma_{k} \Phi_{-\xi_2} = \langle \frac{1}{2} \xi_1 \vert
\sigma_k \vert \frac{1}{2} -\xi_2 \rangle = \sqrt{2} (-1)^{1/2+\xi_2}
\langle \frac{1}{2} \xi_1 \frac{1}{2} \xi_2 \vert 1 k \rangle,
\end{equation}
the $S$ matrix can be written as

\begin{eqnarray}
S & = & i e^2 e_q \frac{1}{(2 \pi)^{1/2}} \frac{m_l}{E_q} \frac{1}{M^2} 2 
\overline{\Psi(0)} \delta^{(4)}(q_1+q_2-P) (-1)^{1/2+\xi_2} \frac{E_q+m_l}{2
m_l} \\
& & \left[ \left(1 + \frac{q^2}{(E_q+m_l)^2} \right) \langle \frac{1}{2}
\xi_1 \frac{1}{2} \xi_2 \vert 1 \mu_J \rangle - \frac{2}{(E_q+m_l)^2} \sum_k
(-1)^k \langle \frac{1}{2} \xi_1 \frac{1}{2} \xi_2 \vert 1 k \rangle q_{-k}
q_{\mu_J} \right].  \nonumber
\end{eqnarray}

Once we have the final expression for $S$, we have to write it in the form 
\cite{weinberg}:

\begin{equation}
S = - 2 \pi i \delta^{(4)}(\sum p_f - \sum p_i) {\cal {M},}
\end{equation}
in order to calculate the differential width of a process with one initial
particle and two final particles as

\begin{equation}
d \Gamma = 2 \pi \vert {\cal M} \vert ^2 \delta^{(4)} (\sum p_f - \sum p_i)
d^3 q_1 d^3 q_2,
\end{equation}
performing an average over initial states and a sum over final states.

In our case,

\begin{eqnarray}  \label{ap-eq}
d \Gamma &=& (2 \pi)^{-2} \delta^{(4)}(q_1+q_2-P) \frac{m_l^2}{E_q^2} \vert 
\overline{\Psi(0)} \vert ^2 \frac{e^4 e_q^2}{M^4} d^3 q_1 d^3 q_2 \left(%
\frac{E_q+m_l}{m_l}\right)^2 \\
& & 3 \frac{1}{3} \sum_{\xi_1 \xi_2 \mu_J} \left[ \left(1 + \frac{q^2}{%
(E_q+m_l)^2} \right) \langle \frac{1}{2} \xi_1 \frac{1}{2} \xi_2 \vert 1
\mu_J \rangle - \frac{2}{(E_q+m_l)^2} \sum_k (-1)^k \langle \frac{1}{2}
\xi_1 \frac{1}{2} \xi_2 \vert 1 k \rangle q_{-k} q_{\mu_J} \right]^2 , 
\nonumber
\end{eqnarray}
where a factor 3 is inferred from the colour part and a factor 1/3 from the $%
2J+1$ initial states average, being $J=1$.

The next step is to compute the square of the modulus of the leptonic part:

\begin{eqnarray}
& & \sum_{\xi_1 \xi_2 \mu_J} \left[ \left(1 + \frac{q^2}{(E_q+m_l)^2}
\right) \langle \frac{1}{2} \xi_1 \frac{1}{2} \xi_2 \vert 1 \mu_J \rangle - 
\frac{2}{(E_q+m_l)^2} \sum_k (-1)^k \langle \frac{1}{2} \xi_1 \frac{1}{2}
\xi_2 \vert 1 k \rangle q_{-k} q_{\mu_J} \right]^2  \nonumber \\
& = & 4 \sum_{\xi_1 \xi_2 \mu_J} \left[ \frac{E_q^2}{(E_q+m_l)^2} \langle 
\frac{1}{2} \xi_1 \frac{1}{2} \xi_2 \vert 1 \mu_J \rangle^2 + \right. 
\nonumber \\
& & - \frac{E_q}{(E_q+m_l)^3} \sum_k (-1)^k \langle \frac{1}{2} \xi_1 \frac{1%
}{2} \xi_2 \vert 1 \mu_J \rangle \langle \frac{1}{2} \xi_1 \frac{1}{2} \xi_2
\vert 1 k \rangle (q_{-k} q_{\mu_J} + q_{-k}^\ast q_{\mu_J}^\ast) + 
\nonumber \\
& & \left. + \frac{1}{(E_q+m_l)^4} \sum_{kl} (-)^{k+l} \langle \frac{1}{2}
\xi_1 \frac{1}{2} \xi_2 \vert 1 k \rangle \langle \frac{1}{2} \xi_1 \frac{1}{%
2} \xi_2 \vert 1 l \rangle q_{-k}^\ast q_{\mu_J}^\ast q_{-l} q_{\mu_J} %
\right]  \nonumber \\
& = & 4 \sum_{\mu_J} \left[\frac{E_q^2}{(E_q+m_l)^2} - \frac{E_q}{(E_q+m_l)^3%
} \sum_k (-1)^k \delta_{\mu_J,k} 2 q_{-k} q_{\mu_J} + \frac{1}{(E_q+m_l)^4}
\sum_{kl} (-)^{k+l} \delta_{kl} q_{-k}^\ast q_{\mu_J}^\ast q_{-l} q_{\mu_J} %
\right]  \nonumber \\
& = & 4 \left[ \frac{3 E_q^2}{(E_q+m_l)^2} - \frac{2 E_q}{(E_q+m_l)^3} q^2 + 
\frac{q^4}{(E_q+m_l)^4} \right]  \nonumber \\
& = & \frac{8 E_q^2}{(E_q+m_l)^2} + \frac{4 m_l^2}{(E_q+m_l)^2}.  \nonumber
\end{eqnarray}

Substituting in Eq. (\ref{ap-eq}),

\begin{equation}
d \Gamma= (2 \pi)^{-2} \delta^{(4)}(q_1+q_2-P) \frac{m_l^2}{E_q^2} \frac{e^4
e_q^2}{M^4} 4 \left( 1 + \frac{2 E_q^2}{m_l^2} \right) \vert \overline{%
\Psi(0)} \vert^2 d^3q_1 d^3q_2.
\end{equation}

One of the two integrals can be done using three of the four Dirac deltas,
from the other one, we obtain a factor $4 \pi$ from the angular part,

\begin{eqnarray}
\Gamma & = & \frac{16 \pi}{4 \pi^2} \frac{m_l^2e^4 e_q^2}{M^4} \int \frac{%
q^2 dq}{q^2 + m_l^2} \delta \left(M - 2 \sqrt{q^2 + m_l^2} \right) \left( 1
+ 2 \frac{q^2 + m_l^2}{m_l^2} \right) \vert \overline{\Psi(0)} \vert^2 
\nonumber \\
& = & \frac{4}{\pi} \frac{m_l^2 e^4 e_q^2}{M^4} \frac{1}{2 \sqrt{1 - 4 \frac{%
m_l^2}{M^2}}} \int \frac{q^2 dq}{q^2 + m_l^2} \delta \left( q - \sqrt{\frac{%
M^2}{4} - m_l^2} \right) \left( 1 + 2 \frac{q^2 + m_l^2}{m_l^2} \right)
\vert \overline{\Psi(0)} \vert^2  \nonumber \\
& = & \frac{2}{\pi} \frac{m_l^2 e^4 e_q^2}{M^4} \frac{1}{2 \sqrt{1 - 4 \frac{%
m_l^2}{M^2}}} \frac{\frac{M^2}{4} - m_l^2}{\frac{M^2}{4}} \left(1 + \frac{M^2%
}{2 m_l^2} \right) \vert \overline{\Psi(0)} \vert^2 ,
\end{eqnarray}
and finally,

\begin{equation}
\Gamma = \frac{16 \pi \alpha^2 e_q^2}{M^2} \left( 1 - 4 \frac{m_l^2}{M^2}
\right)^{1/2} \left(1 + 2 \frac{m_l^2}{M^2} \right) \vert \overline{\Psi(0)}
\vert^2 ,
\end{equation}
which is Eq. (\ref{eq-gam}) without the flavour modification explained in
the text. This procedure is of course more involved and much less elegant
than the trace method, but this is the price to pay for using a
non-relativistic wave function coupled to a good angular momentum inside a
formalism that is covariant at the beginning.

\section{Gaussian electromagnetic density for quarks}

\label{ap-gau}

The density function for quarks must be, in limit of zero size, a Dirac
delta. We can choose a gaussian function:

\begin{equation}
\rho_q(\vec{r}) = \frac{\sigma_q^3}{\pi^{3/2}} e^{-\sigma_q^2 r^2}.
\end{equation}

In the limit $\sigma_q \rightarrow \infty$ the size of the quark tends to
zero and $\rho_q(\vec{r}) \rightarrow \delta^{(3)}(\vec{r})$. The Fourier
transform of the density is

\begin{equation}
\tilde{\rho_q}(\vec{p}) = \frac{1}{(2 \pi)^{3/2}} e^{- \frac{p^2}{4
\sigma_q^2}}.
\end{equation}

Therefore, the quark-antiquark pair density is given by

\begin{eqnarray}
\tilde{\rho_{q\bar{q}}}(\vec{p}) & = & \frac{1}{(2 \pi)^3} e^{- \frac{p^2}{4
\sigma_{q\bar{q}}^2}}, \\
\rho_{q\bar{q}}(\vec{r}) & = & \frac{1}{(2 \pi)^{3/2}} \frac{\sigma_{q\bar{q}%
}^3}{\pi^{3/2}} e^{-\sigma_{q\bar{q}}^2 r^2},
\end{eqnarray}

where

\begin{equation}
\sigma_{q\bar{q}} = \frac{\sigma_q \sigma_{\bar{q}}} {\sqrt{\sigma_q^2 +
\sigma_{\bar{q}}^2}}.
\end{equation}

\begin{table}[tbp]
\caption{Mean value $\langle \protect\chi_{\protect\lambda_1} \vert \protect%
\sigma_i \vert \protect\chi_{\protect\lambda_2} \rangle$ of the Pauli
matrices between helicity states.}
\label{ta-pauli}
\begin{tabular}{ccccc}
$\lambda_1$ & $\lambda_2$ & $\sigma_1$ & $\sigma_0$ & $\sigma_{-1}$ \\ \hline
$1/2$ & $1/2$ & $- \frac{\sin \theta}{\sqrt{2}} e^{i \varphi}$ & $\cos
\theta $ & $\frac{\sin \theta}{\sqrt{2}} e^{-i \varphi}$ \\ 
$1/2$ & $-1/2$ & $- \frac{1 + \cos \theta}{\sqrt{2}} e^{i \varphi}$ & $-
\sin \theta$ & $- \frac{1 - \cos \theta}{\sqrt{2}} e^{-i \varphi}$ \\ 
$-1/2$ & $1/2$ & $\frac{1 - \cos \theta}{\sqrt{2}} e^{i \varphi}$ & $- \sin
\theta$ & $\frac{1 +\cos \theta}{\sqrt{2}} e^{-i \varphi}$ \\ 
$-1/2$ & $-1/2$ & $\frac{\sin \theta}{\sqrt{2}} e^{i \varphi}$ & $- \cos
\theta$ & $- \frac{\sin \theta}{\sqrt{2}} e^{-i \varphi}$%
\end{tabular}
\end{table}

\begin{table}[tbp]
\caption{{\protect\small {Results for the widths with potentials $\protect%
\chi$QM and DNR, in keV. Experimental data are from \protect\cite{pdg}. ``V.
R.'' means application of the original van Royen-Weisskopf formula, ``VRM.''
means the formula taking into account the modifications given by Eq. (\ref
{modif}), and ``dressed'' means that a gaussian distribution have been used
to dress the quarks. The five last transitions ($\ast$) have a theoretical
width $\Gamma_{\protect\mu \protect\mu} \simeq \Gamma_{ee}$, but the
transitions to $\protect\mu^+ \protect\mu^-$ have not been observed
experimentally.}}}
\label{ta-re}
\begin{center}
\begin{tabular}{lccccccc}
& DNR & $\chi$QM & DNR & $\chi$QM & DNR & $\chi$QM &  \\ 
transition & V. R. & V. R. & VRM. & VRM. & dressed & dressed & Exp. \\ \hline
$\rho\rightarrow e^{+} e^{-} $ & 11.360 & 0.998 & 8.601 & 1.315 & 5.719 & 
3.648 & 6.75 $\pm$ 0.33 \\ 
$\ \ \, \rightarrow \mu^{+}\mu^{-}$ & 11.335 & 0.996 & 8.582 & 1.312 & 5.707
& 3.640 & 6.91 $\pm$ 0.42 \\ \hline
$\omega\rightarrow e^{+} e^{-}$ & 1.217 & 0.160 & 0.922 & 0.187 & 0.613 & 
0.460 & 0.60 $\pm$ 0.02 \\ 
$\ \ \, \rightarrow \mu^{+}\mu^{-}$ & 1.215 & 0.160 & 0.920 & 0.187 & 0.612
& 0.459 & $<$1.52 \\ \hline
$\phi\rightarrow e^{+} e^{-}$ & 2.984 & 0.982 & 2.464 & 0.964 & 1.643 & 1.214
& 1.30 $\pm$ 0.03 \\ 
$\ \ \, \rightarrow \mu^{+}\mu^{-}$ & 2.982 & 0.981 & 2.462 & 0.963 & 1.642
& 1.213 & 1.65 $\pm$ 0.22 \\ \hline
$J/\psi$(1S)$\rightarrow e^{+} e^{-}$ & 7.381 & 11.461 & 6.634 & 9.740 & 
5.330 & 4.346 & 5.16 $\pm$ 0.31 \\ 
$\ \ \ \ \ \ \ \ \ \ \ \rightarrow \mu^{+}\mu^{-}$ & 7.381 & 11.461 & 6.634
& 9.740 & 5.330 & 4.346 & 5.11 $\pm$ 0.31 \\ \hline
$\psi$(2S)$\rightarrow e^{+} e^{-}$ & 3.696 & 4.603 & 3.101 & 3.697 & 2.074
& 1.075 & 2.44 $\pm$ 0.45 \\ 
$\ \ \ \ \ \ \ \ \rightarrow \mu^{+}\mu^{-}$ & 3.696 & 4.603 & 3.101 & 3.697
& 2.074 & 1.075 & 2.85 $\pm$ 1.02 \\ 
$\ \ \ \ \ \ \ \ \rightarrow \tau^{+} \tau^{-}$ & 1.436 & 1.788 & 1.204 & 
1.436 & 0.806 & 0.418 & not seen \\ \hline
$\psi$(4040)$\rightarrow e^{+} e^{-}$ & 2.681 & 2.904 & 2.175 & 2.273 & 1.300
& 0.507 & 0.73 $\pm$ 0.25 \\ 
$\ \ \ \ \ \ \ \ \ \ \rightarrow \mu^{+} \mu^{-}$ & 2.681 & 2.904 & 2.175 & 
2.273 & 1.300 & 0.507 & seen \\ 
$\ \ \ \ \ \ \ \ \ \ \rightarrow \tau^{+} \tau^{-}$ & 1.768 & 1.915 & 1.434
& 1.500 & 0.858 & 0.334 & not seen \\ \hline
$\psi$(4415)$\rightarrow e^{+} e^{-}$ & 2.087 & 1.979 & 1.653 & 1.543 & 0.900
& 0.287 & 0.47 $\pm$ 0.24 \\ 
$\ \ \ \ \ \ \ \ \ \ \rightarrow \mu^{+} \mu^{-}$ & 2.087 & 1.979 & 1.653 & 
1.543 & 0.900 & 0.287 & not seen \\ 
$\ \ \ \ \ \ \ \ \ \ \rightarrow \tau^{+} \tau^{-}$ & 1.640 & 1.554 & 1.298
& 1.212 & 0.707 & 0.225 & not seen \\ \hline
$\psi$(3770)$\rightarrow e^{+} e^{-}$ & 0 & 0 & 0.023 & 0.019 & 0.017 & 0.009
& 0.26 $\pm$ 0.05 \\ 
$\ \ \ \ \ \ \ \ \ \ \rightarrow \mu^{+} \mu^{-}$ & 0 & 0 & 0.023 & 0.019 & 
0.017 & 0.009 & not seen \\ 
$\ \ \ \ \ \ \ \ \ \ \rightarrow \tau^{+} \tau^{-}$ & 0 & 0 & 0.011 & 0.009
& 0.008 & 0.004 & not seen \\ \hline
$\psi$(4160)$\rightarrow e^{+} e^{-}$ & 0 & 0 & 0.032 & 0.023 & 0.021 & 0.009
& 0.78 $\pm$ 0.37 \\ 
$\ \ \ \ \ \ \ \ \ \ \rightarrow \mu^{+} \mu^{-}$ & 0 & 0 & 0.032 & 0.023 & 
0.021 & 0.009 & not seen \\ 
$\ \ \ \ \ \ \ \ \ \ \rightarrow \tau^{+} \tau^{-}$ & 0 & 0 & 0.023 & 0.016
& 0.015 & 0.006 & not seen \\ \hline
$\Upsilon$(1S)$\rightarrow e^{+} e^{-}$ & 1.415 & 5.052 & 1.299 & 4.291 & 
1.287 & 2.727 & 1.25 $\pm$ 0.07 \\ 
$\ \ \ \ \ \ \ \ \rightarrow \mu^{+}\mu^{-}$ & 1.415 & 5.052 & 1.299 & 4.291
& 1.287 & 2.727 & 1.30 $\pm$ 0.06 \\ 
$\ \ \ \ \ \ \ \ \rightarrow \tau^{+}\tau^{-}$ & 1.404 & 5.012 & 1.288 & 
4.257 & 1.277 & 2.706 & 1.40 $\pm$ 0.09 \\ \hline
$\Upsilon$(2S)$\rightarrow e^{+} e^{-}$ & 0.654 & 1.496 & 0.588 & 1.252 & 
0.583 & 0.742 & 0.52 $\pm$ 0.12 \\ 
$\ \ \ \ \ \ \ \ \rightarrow \mu^{+}\mu^{-}$ & 0.654 & 1.496 & 0.588 & 1.252
& 0.583 & 0.742 & 0.58 $\pm$ 0.13 \\ 
$\ \ \ \ \ \ \ \ \rightarrow \tau^{+}\tau^{-}$ & 0.650 & 1.487 & 0.585 & 
1.244 & 0.580 & 0.737 & 0.75 $\pm$ 0.71 \\ \hline
$\Upsilon$(3S)$\rightarrow e^{+}e^{-}$ & 0.488 & 0.922 & 0.432 & 0.765 & 
0.428 & 0.446 & seen \\ 
$\ \ \ \ \ \ \ \ \rightarrow\mu^{+}\mu^{-}$ & 0.488 & 0.922 & 0.432 & 0.765
& 0.428 & 0.446 & 0.48 $\pm$ 0.08 \\ 
$\ \ \ \ \ \ \ \ \rightarrow \tau^{+}\tau^{-}$ & 0.485 & 0.917 & 0.430 & 
0.761 & 0.426 & 0.444 & not seen \\ \hline
$\Upsilon$(4S)$\rightarrow e^{+}e^{-}$ & 0.402 & 0.682 & 0.353 & 0.556 & 
0.350 & 0.326 & 0.39 $\pm$ 0.17 \\ 
$\ \ \ \ \ \ \ \ \rightarrow\mu^{+}\mu^{-}$ & 0.402 & 0.682 & 0.353 & 0.556
& 0.350 & 0.326 & not seen \\ 
$\ \ \ \ \ \ \ \ \rightarrow \tau^{+}\tau^{-}$ & 0.400 & 0.679 & 0.352 & 
0.553 & 0.348 & 0.324 & not seen \\ \hline
$\Upsilon$(5S)$\rightarrow e^{+}e^{-}$ & 0.351 & 0.528 & 0.306 & 0.424 & 
0.303 & 0.252 & 0.31 $\pm$ 0.08 \\ 
$\ \ \ \ \ \ \ \ \rightarrow\mu^{+}\mu^{-}$ & 0.351 & 0.528 & 0.306 & 0.424
& 0.303 & 0.252 & not seen \\ 
$\ \ \ \ \ \ \ \ \rightarrow \tau^{+}\tau^{-}$ & 0.349 & 0.526 & 0.305 & 
0.422 & 0.301 & 0.251 & not seen \\ \hline
$\Upsilon$(6S)$\rightarrow e^{+}e^{-}$ & 0.208 & 0.432 & 0.179 & 0.347 & 
0.161 & 0.207 & 0.13 $\pm$ 0.05 \\ 
$\ \ \ \ \ \ \ \ \rightarrow\mu^{+}\mu^{-}$ & 0.208 & 0.432 & 0.179 & 0.347
& 0.161 & 0.207 & not seen \\ 
$\ \ \ \ \ \ \ \ \rightarrow \tau^{+}\tau^{-}$ & 0.207 & 0.430 & 0.178 & 
0.346 & 0.160 & 0.206 & not seen \\ \hline\hline
$\omega(1420)\rightarrow e^{+} e^{-}$ & 0.363 & 0.039 & 0.233 & 0.033 & 0.094
& 0.041 & seen $\ast$ \\ 
$\rho(1450)\rightarrow e^{+} e^{-}$ & 3.063 & 0.202 & 1.970 & 0.192 & 0.794
& 0.326 & seen $\ast$ \\ 
$\omega(1650)\rightarrow e^{+} e^{-}$ & 0 & 0 & 0.021 & 0.012 & 0.011 & 0.009
& seen $\ast$ \\ 
$\phi(1680)\rightarrow e^{+} e^{-}$ & 1.010 & 0.280 & 0.722 & 0.231 & 0.278
& 0.172 & seen $\ast$ \\ 
$\rho(1700)\rightarrow e^{+} e^{-}$ & 0 & 0 & 0.179 & 0.094 & 0.091 & 0.077
& seen $\ast$%
\end{tabular}
\end{center}
\end{table}

\end{document}